\documentstyle[prd,preprint,aps]{revtex}
\begin{document}
\draft

\title{Connections between special relativity, charge conservation, and 
quantum mechanics}
\author{Paul J. Freitas}
\address{Department of Physics, One Shields Avenue, University of
California, Davis, California 95616-8677}
\date{March 24, 1998}

\maketitle

%%%%%%%%%%%%%%%%%%%%%%%%%%% abstract %%%%%%%%%%%%%%%%%%%%%%%%%%%%%%%

\begin{abstract}
Examination of the Einstein energy-momentum relationship suggests that
simple unbound forms of matter exist in a four-dimensional Euclidean space.
Position, momentum, velocity, and other vector quantities can be expressed
as Euclidean four-vectors, with the magnitude of the velocity vector having
a constant value, the speed of light. We see that charge may be simply a
manifestation of momentum in the new fourth direction, which implies that
charge conservation is a form of momentum conservation. The constancy of
speed implies that all elementary free particles can be described in the
same manner as photons, by means of a wave equation. The resulting wave
mechanics (with a few small assumptions) is simply the traditional form of
quantum mechanics. If one begins by assuming the wave nature of matter, it
is shown that special relativistic results follow simply. Thus we see
evidence of a strong connection between relativity and quantum mechanics.
Comparisons between the theory presented here and Kaluza-Klein theories
reveal some similarities, but also many significant differences between
them.
\end{abstract}

\pacs{03.30, 03.65, 03.65.G, 03.65.P}

%%%%%%%%%%%%%%%%%%%%%%%%%% paper contents %%%%%%%%%%%%%%%%%%%%%%%%%

\section{Introduction}
\label{sec:intro}

One of the most important results of twentieth century physics is
Einstein's theory of special relativity. Since its introduction in 1905
\cite{Einstein}, our view of the physical world has been forever changed.
We now have a much better understanding of the relationship between matter
and energy than we ever had before relativity. No physics education is
complete nowadays without a study of special relativity.

Of course, such studies are seldom painless. Although relativity follows
from some simple principles, the logic we use to derive its major results
can be quite complicated. Material objects which possess rest mass must be
treated differently than those that do not in many ways. For example, the
derivation of the transformation for the energy of a particle with rest
mass may involve a lengthy discussion of two-body collisions, as can be
found in Jackson's section 11.5 \cite{Jackson}. The same derivation for a
photon is somewhat simpler, involving only a derivation of the relativistic
Doppler shift and the quantum assumption $E = \hbar \omega$ , as can be
seen in section 11.2 of \cite{Jackson}. Although we seek the transformation
for the same quantity, we find we must use two different methods to find
two different results. It would certainly be desirable if we could treat
both types of matter the same way. It should be noted that there are some
strong similarities between the two formulae. For example, the formula for
the energy of a resting body boosted to a speed $v$ is the same as that of
a photon experiencing a transverse boost of the same magnitude. It would
appear that this is a mere coincidence. Is it?

In this paper we shall see that it is not mere coincidence. It is possible
to add one dimension to our three dimensions of space in such a way that we
can treat simple objects with and without rest mass exactly the same way.
This new four-space has the nice property of being Euclidean, and yields
all of the usual relativistic properties through a few simple, familiar
postulates. By considering a few well-understood physical processes, we
shall see that the momentum of a particle in the fourth direction may
correspond to its charge, which means that charge conservation is just a
form of momentum conservation. We shall also see how easy it is to infer
(with a few small assumptions) that wave and quantum mechanics offer an
excellent description of matter in this four-space. In fact, we can start
with four-dimensional quantum mechanics and derive the results of special
relativity very naturally, with even fewer assumptions. The connection
between relativity and quantum mechanics appears to be stronger than anyone
previously indicated.

Before continuing, let us first define the terminology that we shall use in
this paper. It is standard practice in physics today to refer to the mass
of a resting classical object simply as mass. Here we will instead use the
term rest mass for that quantity, and reserve the term mass for an object's
total relativistic mass. This convention is commonly used in many
elementary treatises on relativity (such as \cite{Resnick&Halliday} and
\cite{French}), and fits in well with the rest mass--relativistic mass
notation found in beginning physics textbooks like \cite{Halliday&Resnick}
and \cite{Tipler}. So here mass is related to total energy by the famous
relationship $E=mc^2$. At the very least, this choice of terminology should
cause no ambiguity, since most readers have probably encountered it before.
Hopefully it will make the discussion more clear as well.

\section{Extension of Space to Four Dimensions}
\label{sec:extension}

Let us first examine a well-known result of special relativity. For any
particle of matter,
\begin{equation}
E^2=p^2c^2+m_0^2c^4. \label{1}
\end{equation}
where $E$, $p$ and $m_0$ are the particle's energy, momentum, and rest
mass, respectively, and $c$ is the speed of light. For reasons which shall
become clear later, we restrict this discussion to extremely simple
particles, such as electrons. Now let us make the following substitutions,
also from special relativity:
\begin{equation}
m_0=m \sqrt{1- {v^2 \over c^2}}
\end{equation}
and
\begin{equation}
\bbox{p}=m\bbox{v}.
\end{equation}
We find that
\begin{equation}
E^2=m^2c^2[v^2+(c^2-v^2)] \label{2}
\end{equation}
Rather than make the obvious simplification to the bracketed term in
(\ref{2}), let us first examine each of its parts separately. If the
particle has the usual Cartesian three-space coordinates $x$, $y$, and $z$,
then
\begin{equation}
v^2=\left(dx \over dt\right)^2+\left(dy \over dt\right)^2+\left(dz \over dt
\right)^2.
\end{equation}
We can make this substitution for the first term in the parentheses of Eq. 
(\ref{2}), but what about the second term? We can find an expression for it
by considering ``proper time'' $t_0$. For an observer in a given reference 
frame, the amount of time $t$ that passes while a time interval $t_0$
passes in a frame moving at speed $v$ (the proper time) is given by the
equation
\begin{equation}
t={t_0 \over \sqrt{1-{v^2 \over c^2}}},
\end{equation}
which implies that
\begin{equation}
t_0=t \sqrt{1-{v^2 \over c^2}}.
\end{equation}
Simple calculus shows that
\begin{equation}
{dt_0 \over dt} = \sqrt{1-{v^2 \over c^2}},
\end{equation}
so
\begin{equation}
c^2\left(dt_0 \over dt\right)^2=c^2-v^2.
\end{equation}
We can now express (\ref{2}) as
\begin{equation}
E^2=m^2c^2\left[\left(dx \over dt\right)^2+\left(dy \over dt\right)^2+
\left(dz \over dt\right)^2+c^2\left(dt_0 \over dt\right)^2\right].
\label{3} \end{equation}
Now we may complete the obvious sum in (\ref{2}).
\begin{equation}
E^2=m^2c^4 \label{4}
\end{equation}
Solving (\ref{3}) and (\ref{4}) simultaneously brings us to an important
result,
\begin{equation}
\left(dx \over dt\right)^2+\left(dy \over dt\right)^2+\left(dz \over
dt\right) ^2+c^2\left(dt_0 \over dt\right)^2=c^2. \label{5}
\end{equation}
Let us now introduce a new quantity, $w$, such that ${dw \over dt}=\pm
c{dt_0 \over dt}$. (We shall see the reason for the sign ambiguity in
section \ref{sec:physical}.) Eq. (\ref{5}) becomes
\begin{equation}
\left(dw \over dt\right)^2+\left(dx \over dt\right)^2+\left(dy \over
dt\right) ^2+\left(dz \over dt\right)^2=c^2. \label{6}
\end{equation}

We have expressed (\ref{6}) in the form of an ordinary self inner product
for a vector in a Euclidean four-space. The last three terms of the sum
were collectively known as $\bbox{v} \cdot \bbox{v}$; now we have an extra
term added. We can incorporate the extra term by simply adding another,
orthogonal direction for ${dw \over dt}$ to our vector space. Let us call
that direction $\bbox{\hat{h}}$, and the directions for ${dx \over dt}$,
${dy \over dt}$, and ${dz \over dt}$ will be the traditional
$\bbox{\hat{\imath}}$, $\bbox{\hat{\jmath}}$ and $\bbox{\hat{k}}$
(respectively) of most elementary physics textbooks. In other words, a
simple particle of matter can be described as traveling through a Euclidean
four-space whose position vector $\bbox{r}$ can be described as
\begin{equation}
\bbox{r}=w\bbox{\hat{h}}+x\bbox{\hat{\imath}}+y\bbox{\hat{\jmath}}+
z\bbox{\hat{k}}. \label{8}
\end{equation}
Its velocity vector ${d\bbox{r} \over dt}$ is constrained by the equation
$|{d\bbox{r} \over dt}|=c$. If we consider instead a photon, whose
three-space speed is $c$, the quantity ${dw \over dt}=0$, so we can use the
same four-space described above. This four-space is sufficiently general to
describe all kinds of matter.

In summary, simple particles of matter can be described as existing in a
Euclidean four-space as described above, with the proper time of the
particle being related to the fourth, non-obvious position component. In
such a space, all simple forms of matter are constantly moving at the speed
of light. Later we shall see how more complicated forms behave.

\section{Physical Interpretations}
\label{sec:physical}

As we saw earlier, ${dw \over dt}$ is related to the rate of change of
proper time for a simple particle, so $|w|$ is, in a classical sense, a
measure of the age of a simple particle. Notice that in Eq. (\ref{6}), we
must allow ${dw \over dt}$ to be either positive or negative. What is the
physical meaning of such a sign?

To answer that question, let us return again to (\ref{3}), slightly
modified: \begin{equation}
E^2=m^2c^2\left[\left(dw \over dt\right)^2+\left(dx \over dt\right)^2+
\left(dy \over dt\right)^2+\left(dz \over dt\right)^2\right]. \label{9}
\end{equation}
The bracketed quantity again appears to be a self inner product of
something fundamental. Let $\bbox{u} \equiv {d\bbox{r} \over dt}$ be our
particle's velocity in our new four-space; (\ref{9}) becomes
\begin{equation}
E^2=(m\bbox{u})^2c^2. \label{10}
\end{equation}
Let us now define the momentum $\bbox{p}$ of a particle in the usual way,
$\bbox{p}=m\bbox{u}$, so
\begin{equation}
E^2=\bbox{p}^2c^2. \label{11}
\end{equation}
Thus all simple matter can be described by the free photon energy-momentum 
relationship $E=pc$. As with the photon, we presume that $E$ is always
positive, which makes perfect sense because $E$ is merely a measure of the 
magnitude of a particle's momentum. Since $E=mc^2$, we should make the same
assumption for $m$.

Now we have a four-dimensional Euclidian momentum space in addition to the 
four-dimensional position space. The momentum space is defined as before,
with the added component
\begin{equation}
\bbox{p}_w=m\left(dw \over dt \right)\bbox{\hat{h}}=\pm m_0c\bbox{\hat{h}}.
\label{12}
\end{equation}
The magnitude of the extra momentum component in these simple particles is 
proportional to their rest mass. Notice the qualifier ``simple.'' Can one
use (\ref{12}) for {\it any} particle? 

Momentum conservation is one of the most useful concepts in physics, with
countless practical applications and experimental verifications. One would 
expect that if rest mass is a form of momentum, then it must conserve in a 
collision. There are, however, many examples of physical processes that
disprove that idea. For example, rest mass does not conserve in an electron
capture process, where a proton and an electron combine to form a neutron.
Thus (\ref{12}) cannot be true for all particles. It does appear to be
correct for simple particles, however. Can we find another physical
quantity that agrees for all particles? We can, and we will now do so by
examining some very simple and well-understood particle interactions.

Let's start with pair annihilation, in which an electron and a positron are
converted to photons. Since neither the electron nor the positron can be
broken down into smaller components, it seems reasonable to assume that
they are among the simplest forms of matter, and that they obey Eq.
(\ref{12}). We expect momentum to conserve here. Since the outgoing photons
are traveling at light speed in three-space, they must each have $p_w=0$.
To ensure momentum conservation, the electron and the positron must have
$w$-momenta equal in magnitude but opposite in sign. There are a few
differences between the electron and the positron, but the fact that they
have opposite charges is the most obvious and important one. It seems very
reasonable to assume that the sign of the charge of a particle is related
to the sign of its $w$-momentum component, and that charge and $w$-momentum
are somehow equivalent.

We can find supporting evidence for this idea by examining the results of
some more complicated interactions, which are all summarized in \cite{PDG}.
Let us assume that an electron has $p_w=-m_ec$ (where $m_e$ is its rest
mass), and a positron has $p_w=m_ec$. A negatively charged muon is known to
decay into an electron and a pair of neutrinos. Neutrinos appear to lack
rest mass \cite{PDG}, so they must also lack $w$-momentum. We expect
$w$-momentum to conserve, so the electron and the muon must have the same
$w$-momentum. Since the negative tauon can also decay to an electron and
two neutrinos, it must also have $p_w=-m_ec$. A negative tauon can also
decay to a negative pion and a neutrino, so the negative pion must also
have $p_w=-m_ec$. We can conclude that electrons and negative muons,
tauons, and pions all have the same $w$-momentum. It appears that the
$w$-momentum of a particle depends only on its charge $q$. We can express
this idea mathematically as 
\begin{equation}
p_w=-{qm_ec \over e}, \label{12a}
\end{equation}
where $e$ is the charge of an electron. It is well-known that charge
conserves; we can now see that charge conservation may be simply another
form of momentum conservation.

\section{Wave Mechanics and Quantum Mechanics}
\label{sec:wave}

Earlier we observed that, with the inclusion of a form of proper time as a 
spatial coordinate, simple free particles can be viewed as traveling at the
speed of light at all times. This property of constant velocity strongly
suggests that simple free particles can be described mathematically in the
same manner as photons; they are simply solutions to some form of the wave
equation. In this section we shall explore this idea and examine its
consequences.

Let $\bbox{\Psi}(\bbox{x},t)$ be the wave function for a free particle at
position $\bbox{x}$ and time $t$. The dimensionality of $\bbox{\Psi}$ is
not important for this discussion at this time. Also, let $x_j \ (j =
1,2,3,4)$ be a set of Cartesian coordinates that span our four-dimensional
space, with $w=x_4$. Our wave equation is:
\begin{equation}
\sum^{4}_{j=1}{\partial ^2 \bbox{\Psi} \over \partial x_j^2}={1 \over c^2} 
{\partial ^2 \bbox{\Psi} \over \partial t^2}. \label{13}
\end{equation}
Solutions to this equation are well-known. In these coordinates, we can say
that the eigenfunction $\bbox{\Psi}$ for wave vector $\bbox{k}$ is
\begin{equation}
\bbox{\Psi}(\bbox{x},t)=\bbox{\Psi_0}e^{i(\bbox{k} \cdot \bbox{x} - \omega 
t)}. \label{14}
\end{equation}
where $\omega ={|\bbox{k}| \over c}$. To turn this wave mechanical system
into a form of quantum mechanics, we need only one extra assumption. We
must simply relate the momentum and wave vectors through the equation
$\bbox{p}=\hbar \bbox{k}$ and our wave mechanics becomes the quantum
mechanics of free photons.

One useful result of quantum mechanics is the operator formalism. The forms
of the operators often seem cryptic, but one nice feature of the wave
mechanics we are using is that the operators follow rather intuitively. To 
find them, all we need to do is take our eigenfunctions and ask what we
need to do to them to find the properties which we are interested in. For
example, simple calculus tells us that, for a function of the form of
(\ref{14}),
\begin{equation}
-i \hbar {{\partial \bf \Psi}\over \partial x_j}=p_j\bbox{\Psi}.\label{15} 
\end{equation}
We could mathematically extract $p_j$ from this equation, but we know that 
the operator form is useful in and of itself. We can use Eq. (\ref{15}) as
a definition for momentum in more complicated systems, by defining the
momentum operator $\hat{P}_j$ as
\begin{equation}
\hat{P}_j=-i \hbar {\partial \over \partial x_i}.\label{16}
\end{equation}
This result is simply the operator notation found in most quantum mechanics
textbooks (for example, see \cite{Shankar} and \cite{Schiff}). The
arguments that once led to these operators were traditionally quite
complicated, so much so that most introductory textbooks will not discuss
them, but now we have a more intuitive explanation. Using the same
reasoning, one can also define the quantum mechanical energy operator
$\hat{E}$ as
\begin{equation}
\hat{E}=i\hbar{\partial \over \partial t}.\label{17}
\end{equation}
By simply multiplying (\ref{13}) by $-\hbar^2c^2$ we see that the expected 
relation
\begin{equation}
\hat{E}^2\bbox{\Psi}=\hat{P}^2c^2\bbox{\Psi} \label{18}
\end{equation}
holds true.

Equation (\ref{18}) appears to be a quantum mechanical equation for free
photons, assuming that $\bbox{\Psi}$ is a scalar, $\Psi$. How can we turn
this into a corresponding relationship for free particles with rest mass?
Let us start by substituting in a form of (\ref{12}), namely
\begin{equation}
p_4 \Psi = -i \hbar {\partial \Psi \over \partial x_4} = \pm m_0c \Psi
\end{equation}
and (\ref{18}) becomes
\begin{equation}
-\hbar^2{\partial^2 \Psi \over \partial t^2}=-\hbar^2c^2\nabla^2\Psi+
(m_0c^2)^2\Psi.
\end{equation}
Rearranging and dividing by $\hbar^2c^2$ gives
\begin{equation}
-{1 \over c^2}{\partial ^2 \Psi \over \partial t^2}+\nabla ^2
\Psi=\left({m_0c \over \hbar}\right)^2 \Psi, \label{21}
\end{equation}
which is of course the Klein-Gordon equation. It is now a simple matter to 
reduce (\ref{21}) to the free particle Schr\"odinger equation.

Thus we can infer from some simple relativistic considerations that quantum
mechanics is applicable in our four-space. With only a few fairly intuitive
assumptions, one can derive for free particles the quantum mechanical wave
equation, the Klein-Gordon equation, and the Schr\"odinger equation.
Without the addition of a fourth orthogonal dimension to our definition of
space, quantum mechanics and relativity are independent theories. In our
Cartesian four-space, however, we can see that there is in fact a strong
interdependence between them.

\section{Rest Mass in Physical Systems}
\label{sec:rest}

Earlier we established that the fourth momentum component of a particle was
not necessarily related to its rest mass. In this section we shall take a
closer look at rest mass, and find out what it is and why relates to mass
in the standard relativistic sense. We shall use quantum mechanics as the
tool for our investigation.

Imagine two free particles overlapping in space in the center-of-mass
reference frame. Let them have wave vectors $\bbox{-k}$ and $\bbox{k}$
respectively. Their wavefunctions are
\begin{equation}
\bbox{\Psi_1}(\bbox{x},t)=\bbox{\Psi_0}e^{i(\bbox{-k} \cdot \bbox{x}
-\omega t)}
\end{equation}
and
\begin{equation}
\bbox{\Psi_2}(\bbox{x},t)=\bbox{\Psi_0}e^{i(\bbox{k} \cdot \bbox{x} -\omega
t)}.
\end{equation}
From the operators defined in section \ref{sec:wave} we can see that the
momentum for the combination must be 0, but the energy $E$ is equal to $2
\hbar |\bbox{k}|$. The combination is in a zero momentum state, but it has
a significant amount of energy. If we use the standard relativistic
definition of mass ($E=mc^2$), we can say that the rest mass of this system
is $2 \hbar |\bbox{k}| \over c^2$.

So how does the energy of this system transform after a velocity shift
$\bbox{v}$? We can think of the two wave equations as being the same as
those for photons with opposite momenta. The formula for the energy
transformation can be determined from adding Doppler shift results for each
wave function, which are
\begin{equation}
E' = {{E - \bbox{v} \cdot \hbar \bbox{k}} \over \sqrt{1 - {v^2 \over
c^2}}}.
\end{equation}
Because we are starting with two waves of opposing momenta, the final
result, independent of the direction of the velocity shift, is
\begin{equation}
E' = {E \over \sqrt{1 - {v^2 \over c^2}}} \label{23}
\end{equation}
so in the transformed reference frame, if $E'$ is the transformed energy
and $E$ is the untransformed energy, we can use the relationships
$E=\hbar\omega$ and $E=mc^2$ to find that
\begin{equation}
m = {m_0 \over \sqrt{1 - {v^2 \over c^2}}}, \label{24}
\end{equation}
where $m_0$ is a constant. It can now be easily shown that for this
multiparticle system, mass obeys Eq. (\ref{1}), regardless of the direction
of the velocity change. It is interesting to notice that there is no
$w$-momentum in this system, but there is rest mass. Rest mass is not
necessarily, then, a phenomenon of momentum in systems composed of several 
particles.

For systems composed of more than two particles, the same results apply,
which can be seen by simply pairing off particles with opposing momenta. If
an excess of $w$-momentum exists in the system, we will not notice any
strange effects, because all of our velocity shifts are done in
three-space. Rest mass and $w$-momentum are both invariant.

We can now see that $\bbox{p}$ is a four-vector, but in many cases of
interest, one can simply use the three-vector form instead. For example, it
is currently common practice to simply assume that electrons and positrons 
have rest mass instead of $w$-momentum; this approximation works well in
many circumstances. For compound systems which are neutral, there is no
$w$-momentum, so the three-vector solution is exact. For charged compound
systems, since we do all of our velocity shifts in three-space, rest mass
and $w$-momentum are both invariant. One could treat ${p_w}^2c^2 +
{m_0}^2c^4$ in Eq. (\ref{1}) as a single rest mass term; there are no
differences at this time.

\section{Derivations of Special Relativistic Phenomena}
\label{sec:derivations}

Let us now look back at our four-dimensional quantum mechanics. When we
derived its properties in section \ref{sec:wave}, we started from
relativistic considerations and developed quantum mechanics after making a 
few assumptions. Here we shall see that if we start from a four-dimensional
quantum mechanical perspective, we can derive the key results from special 
relativity very easily.

As a starting point, we need an expression for quantum mechanics. Since
special relativity is a theory for free particles, let us assume that there
are no interactions. Our quantum mechanical wave equation is simply
equation (\ref{13}), with the eigenfunctions as shown in equation
(\ref{14}).

We should first derive the Lorentz transformations. A common way to derive 
them involves the {\it gedankenexperiment} of examining the travel of a
burst of light propagating in all directions from the perspective of two
different reference frames, as Einstein first did \cite{Einstein}. We can
derive these transformations in exactly the same way in our
four-dimensional space. There is an added direction, which turns Einsteinþs
three-dimensional sphere of light into a four-dimensional sphere described
by the equation
\begin{equation}
w^2+x^2+y^2+z^2=c^2t^2 \label{26}
\end{equation}
but there are no other differences in the proof. The $w$-direction is
simply another direction orthogonal to a special relativistic velocity
shift, so the transformation of $w$ is simple, $w'=w$. So distance in the
$w$-direction is invariant in special relativity. This is not new
information, however. Let us rewrite Eq. (\ref{26}) in a differential form
($x \rightarrow dx$, etc.) and solve for $dw^2$:
\begin{equation}
dw^2=c^2dt^2-dx^2-dy^2-dz^2 \label{26a}
\end{equation}
We can now see $dw$ for what it truly is; it is the definition of what is
commonly referred to as the spacetime interval, which was first discussed
by Minkowski \cite{Minkowski}. The spacetime interval is simply another
distance for a direction orthogonal to the direction of the velocity shift,
and now has a fairly simple explanation.

Now that we have the Lorentz transformations, the energy and momentum
transformations follow in a simple manner. A free particle is described as
a combination of eigenfunctions of the form of Eq. (\ref{14}). We can
examine each eigenfunction individually. Observers in all reference frames
must agree on the phase of an eigenfunction, for exactly the same reason
that observers in different frames must agree on the phase of a photon's
wavefunction. In other words, by adding an extra dimension to space, we can
treat photons and particles with rest mass in exactly the same way. For
photons, we can derive the relativistic Doppler shift equations to find the
energy and momentum transformations (see section 11.2 of \cite{Jackson});
these same formulae apply to other kinds of matter as well in a quantum
mechanical four-space. There is absolutely no difference in the way these
types of matter are treated.

Thus we have shown that quantum mechanics and special relativity are very
closely related. All of the properties of matter in special relativity are 
simple consequences of the four-dimensional quantum mechanics we are
considering. It is interesting to note that relativistic properties
followed very simply from quantum mechanics, whereas we needed to make
several assumptions (albeit simple ones) to derive our quantum mechanics
from relativistic considerations. Instead of saying that our quantum
mechanics is relativistic, perhaps it would be better to say that special
relativity is quantum mechanical.

\section{Comparison With Kaluza-Klein Theories}
\label{sec:Kaluza-Klein}

Those readers who are familiar with the general theory of relativity may
see some similarities between the ideas proposed here and Kaluza-Klein
theories \cite{KKTheories}. These theories in their basic form have a
number of similarities to the idea proposed above, such as the addition of
another dimension of space to explain electric charge. In this section we
shall discuss the similarities between my theory and Kaluza-Klein theories.
We shall see that although they share some common characteristics, they are
in fact different in many respects, enough so that they cannot be
considered the same theory.

Let us begin with a very brief description of Kaluza-Klein theories. A
Kaluza-Klein space consists of a four-dimensional space as described by
general relativity (three dimensions of space plus time) plus a fifth,
spacelike dimension.  The fifth dimension of Kaluza-Klein theories is used
in a very similar way to the theory presented above, to explain the
electric charge of an object. The fifth dimension can be thought of as a
small circle with circumference $l$, which leads to quantization of
electric charge.

Now let us compare the two theories. In the theory presented above, no
condition of circularity was imposed on the $w$-dimension; space is assumed
to be infinite in all dimensions, which is a large difference between the
theory presented above and Kaluza-Klein theories. My theory is different in
other respects as well. As we saw above, my theory has its origins in
special relativity, so space is flat. Kaluza-Klein theories have their
roots in general relativity, and their spaces are curved and closed in at
least one dimension. Also, the two theories serve different purposes.
Kaluza-Klein theories were originally created in an attempt to unite
fundamental forces (such as electromagnetism and gravitation) in a general
relativistic theory. My theory demonstrates a different kind of
unification, between quantum mechanics and special relativity.

The different natures of these two theories leads to very different
physical predictions. Although in this paper we did not make the
$w$-dimension periodic, we could do so in order to quantize charge. There
may be other explanations for charge quantization, but let us assume the
$w$-dimension is periodic here purely for the sake of argument. One might
argue that by doing so, and by assuming that the Kaluza-Klein spacetime
lacks gravitation, Kaluza-Klein theory and the theory presented here are
one and the same. This statement is incorrect. The circumference of
Kaluza-Klein theory was calculated by Klein \cite{KleinNature} to be
\begin{equation}
l = { {hc \sqrt{2 \kappa} } \over e}=8 \times 10^{-33} \rm{m},
\end{equation}
where $\kappa$ is the Einstein gravitational constant. In contrast, the
theory presented in this paper indicates that $l$ is the wavelength of an
electron with no three-space momentum. This wavelength is commonly known as
the Compton wavelength, so
\begin{equation}
l = {h \over {m_e c}} =2.426 \times 10^{-12} \rm{m},
\end{equation}
which is many orders of magnitude larger than Klein's result.

In summary, the two theories are in fact quite different from one another, 
despite their similar uses of extra dimensions. The theories have different
origins; Kaluza-Klein theories derive from general relativity, while my
theory derives from special relativity and empirical observation. The
theories attempt to unify completely different aspects of physics;
Kaluza-Klein theories attempt to unify physical forces, while my theory
attempts to unify different physical pictures, those of quantum mechanics
and relativity. The shapes of the spaces are quite different; Kaluza-Klein
theories use curved spaces closed in at least one dimension, while the
theory described here uses an infinite flat space. Even if we make an
effort to shape the two spaces similarly, as we did above, we find we are
forced to size them differently.

\section{Conclusion}
\label{sec:conclusions}

By defining a fourth component in position space related to the
relativistic proper time and the spacetime interval, we can describe the
velocity of any elementary particle of matter as having the magnitude of
the speed of light. The resulting four-space is Euclidean, which is a nice
consequence in and of itself. A corresponding fourth momentum component,
related to the charge and rest mass of the elementary particle, can also be
defined. As a consequence, any elementary free particle can be treated
exactly the same way, regardless of rest mass. Thus we can unify our
description of free particles to a large extent.

There are several interesting consequences of this new four-space. For
example, conservation of charge may simply be a restatement of $w$-momentum
conservation. Also, the constant velocity condition strongly suggests that 
matter is obeying a wave equation, as photons do, so quantum mechanics is
in fact a very natural way to describe physical phenomena in this
four-space. Also, one can work backwards and show that relativistic
consequences result in a simple way from using this quantum mechanical
system. It took some intuition to derive quantum mechanics from relativity,
but relativity is a simple consequence of quantum mechanics. One may still
argue about which is more fundamental, relativity or quantum mechanics, but
this argument has no meaning. The two theories cannot be neatly separated
anymore. We can now see that they are like two different facets of the same
gem.

%%%%%%%%%%%%%%%%%%%%%%% references PRD style %%%%%%%%%%%%%%%%%%%%%%%%%%%%%


\begin{references}

\bibitem{Einstein}A. Einstein, Ann. Phys. (Leipzig) {\bf 17}, 1905. [A.
Einstein, H.A. Lorentz, H. Weyl, and H. Minkowski, {\it The Principle of
Relativity: A Collection of Papers on the Special and General Theory of
Relativity}, notes by A. Sommerfeld (Dover, New York, 1952), p. 35.]
\bibitem{Jackson}J.D. Jackson, {\it Classical Electrodynamics}, Second
Edition (Wiley, New York, 1975).
\bibitem{Halliday&Resnick}D. Halliday and R. Resnick, {\it Fundamentals of 
Physics}, Third Edition (New York, Wiley, 1988).
\bibitem{Resnick&Halliday}R. Resnick and D. Halliday, {\it Basic Concepts
in Relativity and Early Quantum Theory}, Second Edition (New York, Wiley,
1985). \bibitem{Tipler}F. Tipler, {\it Physics for Scientists and
Engineers}, Third Edition (New York, Plenum, 1994).
\bibitem{French}A.P. French, {\it Special Relativity} (New York, Norton,
1968).
\bibitem{PDG}Particle Data Group, Phys. Lett. B {\bf 239} (1990).
\bibitem{Shankar}R. Shankar, {\it Principles of Quantum Mechanics}, Second 
Edition (New York, Plenum, 1994).
\bibitem{Schiff}L. Schiff, {\it Quantum Mechanics}, Third Edition (New
York, McGraw-Hill, 1968).
\bibitem{Minkowski}H. Minkowski, in {\it The Principle of Relativity},
(Dover, New York, 1952), p. 73.
\bibitem{KKTheories}{\it Modern Kaluza-Klein Theories}, edited by T.
Appelquist, A. Chodos, and G. O. Freund (Addison-Wesley, Menlo Park, 1987).
\bibitem{Kaluza}T. Kaluza, Sitzungsber. Preuss. Akad. Wiss. Phys. Math.
Klasse (1921), 966. [T. Kaluza, in {\it Modern Kaluza-Klein Theories}, p.
61].
\bibitem{KleinZF}O. Klein, Z. Phys. {\bf 37}, 895 (1926). [O. Klein, in
{\it Modern Kaluza-Klein Theories}, p. 76.]
\bibitem{KleinNature}O. Klein, Nature {\bf 118}, 516 (1926). [O. Klein, in 
{\it Modern Kaluza-Klein Theories}, p. 88.]

\end{references}
\end{document}